\documentclass[reqno]{amsart}
\usepackage{amsthm}

\newcommand\R{\mathbb R}
\newcommand\C{\mathbb C}

\newcommand\cO{\mathbb O}
\newcommand\x{\times}

\theoremstyle{plain}
\newtheorem{thm}{Theorem}

\newtheorem{lem}[thm]{Lemma}

\theoremstyle{remark}
\newtheorem*{remark}{Remark}

\begin{document}

\title[Boundary conditions]{Boundary 
conditions for the Einstein-Christoffel formulation of Einstein's equations}
\author{Douglas N.~Arnold}
 \email{arnold@ima.umn.edu}
\urladdr{http://www.ima.umn.edu/\char'176arnold}
\address{Institute for Mathematics and its Applications, University
of Minnesota, Minneapolis, Minnesota 55455.}
\author{Nicolae Tarfulea}
 \email{tarfulea@purdue.calumet.edu}
\urladdr{http://ems.calumet.purdue.edu/tarfulea}
\address{Department of Mathematics, Purdue University Calumet, Hammond,
Indiana 46323.}

\date{}


\begin{abstract}
Specifying boundary conditions continues to be a challenge in numerical 
relativity in order to obtain a long time convergent numerical simulation of
Einstein's equations in domains with artificial boundaries.
In this paper, we address this problem for the Einstein--Christoffel (EC) symmetric
hyperbolic formulation of Einstein's equations linearized
around flat spacetime. First, we prescribe simple boundary conditions that make the
problem well posed and preserve the constraints. Next, we indicate
boundary conditions for a system that extends the linearized EC system by including the
momentum constraints and whose solution solves Einstein's equations in a bounded domain.
\end{abstract} 

\maketitle

\section{Introduction}
In the Arnowitt--Deser--Misner or ADM decomposition, Einstein's equations split into a 
set of evolution equations and a set of constraint
equations (see Section~\ref{ADM_dec}), and what one does to construct a solution consists of
first specifying the initial data that satisfies the constraints
and then applying the evolution equations to compute the solution
for later times. The problem of well-posedness in the analytic sense
has been intensely studied, with the result that there is a great
deal of choice of formulations available for analytic
studies (see \cite{BS}, \cite{SN}, \cite{FR}, \cite{AY}, \cite{KST}, \cite{AA}, \cite{Ar0}, 
\cite{Ar}, \cite{AACBY}, \cite{As1}, 
\cite{BM}, \cite{BMSS}, \cite{CB2}, \cite{CBR}, \cite{FRe}, \cite{FG1}, \cite{Re},
\cite{Re2}, among others). 
However, in numerical relativity, one
usually solves the Einstein equations in a bounded domain (cubic
boxes are commonly used) and the question that arises is what
boundary conditions to provide at the artificial boundary. 
In general, most numerical approaches have been made using carefully 
chosen initial data that satisfies the constraints. On the other hand, finding appropriate boundary
conditions that lead to well-posedness and consistent with constraints
is a difficult problem and subject to intense investigations in the recent years.
In 1998, Stewart \cite{S} has addressed this subject within Frittelli--Reula formulation \cite{FR}
linearized around flat space with unit lapse and zero shift in the quarter plane. Both
main system and constraints propagate as first order strongly hyperbolic systems. 
This implies that vanishing values of the constraints at $t=0$ will propagate along characteristics.
One wants the values of the incoming constraints at the boundary to vanish. However, one can not just impose
them to vanish along the boundaries since the constraints involve derivatives of the fields
across the boundary, not just the values of the fields themselves. If the Laplace--Fourier
transforms are used, the linearity of the differential equations gives algebraic equations
for the transforms of the fields. Stewart deduces boundary conditions for the main
system in terms of Laplace--Fourier transforms that preserve the constraints by imposing the
incoming modes for the system of constraints to vanish and translating these conditions in terms of 
Laplace--Fourier transforms of the main system variables.
In 1999, a well posed initial-boundary value formulation was given by Friedrich and Nagy \cite{FN}
in terms of a tetrad-based Einstein--Bianchi formulation.
In view of our work which is to be presented here, of particular interest 
are the more recent investigations regarding special boundary conditions that
prevent the influx of constraint violating modes into the computational
domain for various hyperbolic formulations of Einstein's equations
(see \cite{Al}, \cite{Al2}, \cite{AM}, \cite{BB}, \cite{CLT}, \cite{CPRST}, \cite{KLSBP},
\cite{HLOPSK}, \cite{Sa}, \cite{Sc}, \cite{ST}, \cite{Ta}, \cite{Ta2}, among others). 
A different approach can be found in \cite{FG2}, where the authors stray away from
the general trend of seeking to impose the constraints along the boundary. They argue
that the projection of the Einstein equations along the normal to the boundary yields
necessary and appropriate boundary conditions for a wide class of equivalent formulations.
The ideas and techniques introduced in \cite{FG2} are further developed and proven to be
effective by the same authors in \cite{FG0}. In principal, they show that the projection of the
Einstein tensor along the normal to the boundary relates to the propagation of the
constraints for two representations of Einstein's equations with vanishing shift vector, 
namely, the Arnowitt--Deser--Misner (ADM) formulation \cite{ADM} and the classical
Einstein-Christoffel (EC) formulation \cite{AY}. In particular, for the EC formulation
they retrieve a subclass of the boundary conditions presented in \cite{CPRST} and
\cite{AT}.

Of course, specifying constraint-preserving boundary conditions for a 
certain formulation of Einstein's equations does not solve entirely the 
complicated problem of numerical relativity. There are other aspects that have to be
addressed in order to obtain good numerical simulations; for example, the
existence of  bulk constraint
violations, in which existing violations are amplified by the evolution equations (see 
\cite{CHLP}, \cite{CHLP2}, \cite{LSKPST}, \cite{Sch}, and references therein).
A review of some work done in this direction can be found in the introductory
section of \cite{KLSBP}.
Before we end this very brief review, it should also be mentioned the work done on boundary
conditions for Einstein's equations in harmonic coordinates, when Einstein's equations
become a system of second order hyperbolic equations for the metric components.
The question of the constraints preservation does not appear here, as it is hidden
in the gauge choice, i.e., the constraints have to be satisfied only at the initial
surface, the harmonic gauge guarantees their preservation in time 
(see \cite{Pr}, \cite{SSW}, \cite{SW}, and references therein).

In this paper we address the boundary conditions problem for the classical 
EC equations derived in \cite{AY}, linearized with respect to
the flat Minkowski spacetime, and with arbitrary lapse density and
shift perturbations. 
This problem has been addressed before in \cite{CLT} in the case of spherically symmetric
black-hole spacetimes in vacuum or with a minimally coupled scalar field, within the
EC formulation of Einstein's equations. Here Stewart's idea of imposing the
vanishing of the ingoing constraint modes as boundary conditions is employed
once again. Then, the radial derivative is eliminated in favour of the time derivative
in the expression of the ingoing constraints by using the main evolution system.
The emerging set of boundary conditions depends only on the main variables
and their time derivative and preserves the constraints. In \cite{CPRST} this technique is
refined and employed for the generalized EC formulation \cite{KST} when
linearized around Minkowski spacetime with vanishing lapse and shift perturbations
on a cubic box. Again, the procedure consists in choosing well-posed boundary
conditions to the evolution system for the constraint variables and translating
them into well-posed boundary conditions for the variables of the main evolution system. 
The scheme proposed in \cite{CPRST} ends up giving
two sets, called ``Dirichlet and Neumann-like,'' of constraint preserving boundary conditions.
However, the energy method used in \cite{CPRST} works only for symmetric hyperbolic
constraint propagation, which forces the parameter $\eta$ of the generalized EC system
to satisfy the condition $0<\eta<2$. Therefore the analysis in \cite{CPRST} does
not cover the case $\eta=4$ required for the standard EC formulation introduced in \cite{AY}.
In \cite{AT} we announced and presented our results on the boundary conditions problem for 
the standard EC formulation ($\eta=4$) linearized around the Minkowski spacetime with
arbitrary lapse density and shift perturbations in the
Penn State Numerical Relativity Seminar. In essence, we introduced the very same 
sets of boundary conditions that are under scrutiny in this material,
i.e., \eqref{set2b} and \eqref{set1b} (see \cite{AT}). 
Much of this material appeared also in the thesis of the second author \cite{Ta}.

The organization of this paper is as follows: in Section~\ref{ADM_dec} we
introduce Einstein's equations and their ADM equations for vacuum spacetime. 
In Section~\ref{LEC}, by densitizing
the lapse, linearizing, and defining a set
of new variables, we derive the linearized EC first order symmetric
hyperbolic formulation around flat spacetime. The equivalence of
this formulation with the linearized ADM is proven in the Cauchy
problem case. In Section~\ref{MNCPBC} we indicate two distinct sets of
well-posed constraint-preserving boundary conditions for the
linearized EC. We prove that the linearized EC together with
these boundary conditions is equivalent with linearized ADM on
polyhedral domains. In Section~\ref{EECS} we indicate
boundary conditions for an extended unconstrained system equivalent to the
linearized ADM decomposition.
In Section~\ref{IBC} we discuss the case of inhomogeneous boundary conditions.
We end this work with a summary and a discussion of our results in Section~\ref{CR}. 
For reader's convenience, in the appendix we review a classical result on 
the $L^2$ well-posedness of maximal nonnegative boundary conditions for 
symmetric hyperbolic systems. 

\section{Einstein's Equations and the ADM Decomposition}
\label{ADM_dec}
In general relativity, spacetime is a 4-dimensional manifold $M$ of events endowed with
a pseudo-Riemannian metric $g_{\alpha\beta}$ that determines the length of the line element
$ds^2=g_{\alpha \beta}dx^{\alpha}dx^{\beta}$. This metric determines curvature on the manifold, 
and Einstein's equations relate the curvature at a point of spacetime to the
mass-energy there: $G_{\alpha \beta}=8\pi T_{\alpha \beta}$,
where $G_{\alpha \beta }$ is the {\it Einstein tensor}, i.e., the {\it trace-reversed} Ricci tensor
$G_{\alpha\beta}:=R_{\alpha\beta}-\frac{1}{2}Rg_{\alpha\beta}$, and $T_{\alpha \beta }$
is the {\it energy-momentum tensor}. In what follows we will restrict ourselfs to the case of
vacuum spacetime, that is $T_{\alpha\beta}=0$.
Einstein's equations can be viewed as equations for geometries, that is, their solutions are
equivalent classes under spacetime diffeomorphisms of metric tensors. To break this diffeomorphisms
invariance, Einstein's equations must be first transformed into a system having a well-posed Cauchy
problem. In other words, the spacetime is foliated and each slice $\Sigma_t$ is characterized by its
intrinsic geometry $\gamma_{ij}$ and extrinsic curvature $K_{ij}$, which is essentially the
``velocity'' of $\gamma_{ij}$ in the unit normal direction to the slice. Subsequent slices are connected
via the lapse function $N$ and shift vector $\beta^i$ corresponding to the ADM decomposition \cite{ADM} 
(also \cite{Y2}) of the line element
\begin{equation}
ds^2=-N^2dt^2+\gamma_{ij}(dx^i+\beta^idt)(dx^j+\beta^jdt).
\end{equation}
This decomposition allows one to express six of the ten components of Einstein's equations in vacuum as 
a constrained system of evolution equations for the metric $\gamma_{ij}$ and the extrinsic curvature $K_{ij}$:
\begin{equation}\label{ADM}
\begin{gathered}
\dot{\gamma}_{ij} = -2NK_{ij} + 2 \nabla_{(i}\beta_{j)},
\\
\dot{K}_{ij} =
N[R_{ij}+(K_l^l)K_{ij}-2K_{il}K_j^l]+\beta^l\nabla_l K_{ij}
+K_{il}\nabla_j\beta^l + K_{lj}\nabla_i\beta^l-\nabla_i\nabla_j N,
\\
R_i^i + (K_i^i)^2-K_{ij}K^{ij}=0,
\\
\nabla^jK_{ij}-\nabla_iK_j^j=0.
\end{gathered}
\end{equation}
where we use a dot to denote time differentiation and $\nabla_j$ for the covariant derivative
associated to $\gamma_{ij}$. The spatial Ricci tensor
$R_{ij}$ has components given by second order spatial differential operators applied to
the spatial metric components $\gamma_{ij}$. 
Indices are raised and traces taken with respect to the spatial metric $\gamma_{ij}$, and paranthesized
indices are used to denote the symmetric part of a tensor.

\section{Linearized Einstein--Christoffel}
\label{LEC}
The Einstein--Christoffel or EC formulation \cite{AY} is derived from the ADM system
with a \emph{densitized lapse}.  That is, we replace the lapse $N$
in \eqref{ADM} with
$\alpha\sqrt g$ where $\alpha$ denotes the lapse density.  A trivial
solution to this system is Minkowski spacetime in Cartesian
coordinates, given by $\gamma_{ij}=\delta_{ij}$, $K_{ij}=0$, $\beta^i=0$,
$\alpha=1$. In the remainder of the paper we will consider the problem
linearized about this solution.  To derive the linearization, we write
$\gamma_{ij}=\delta_{ij}+\bar g_{ij}$, $K_{ij}=\bar K_{ij}$,
$\beta^i=\bar\beta^i$, $\alpha=1+\bar\alpha$, where the bars indicate
perturbations, assumed to be small.  If we substitute these expressions
into \eqref{ADM} (with $N=\alpha\sqrt \gamma$), and ignore terms which
are at least quadratic in the perturbations and their derivatives, then
we obtain a linear system for the perturbations. Dropping the bars, the
system is
\begin{gather}
\label{G}
\dot g_{ij}=-2K_{ij}+2\partial_{(i}\beta_{j)},
\\ \label{KADM}
\dot K_{ij}=\partial^l\partial_{(j}g_{i)l}-\frac12\partial^l\partial_l
g_{ij}-\partial_i\partial_jg_l^l-\partial_i\partial_j\alpha,
\\ \label{C}
C:=\partial^j(\partial^lg_{lj}-\partial_jg_l^l)=0,
\\ \label{Cj}
C_j:=\partial^lK_{lj}-\partial_jK_l^l=0,
\end{gather}
where we use a dot to denote time differentiation.
\begin{remark}
For the linear system the effect of densitizing the lapse is to change
the coefficient of the term $\partial_i\partial_j g_l^l$ in
\eqref{KADM}.  Had we not densitized, the coefficient would have
been $-1/2$ instead of $-1$, and the derivation of the linearized
EC formulation below would not be possible.
\end{remark}

The usual approach to solving the system \eqref{G}--\eqref{Cj} is to
begin with initial data $g_{ij}(0)$ and $K_{ij}(0)$ defined on $\R^3$
and satisfying the constraint equations \eqref{C}, \eqref{Cj}, and to
define $g_{ij}$ and $K_{ij}$ for $t>0$ via the Cauchy problem for the
evolution equations \eqref{G}, \eqref{KADM}.  It can be easily shown
that the constraints are then satisfied for all times.  Indeed, if we
apply the Hamiltonian constraint operator defined in \eqref{C} to the
evolution equation \eqref{G} and apply the momentum constraint operator
defined in \eqref{Cj} to the evolution equation \eqref{KADM}, we obtain
the first order symmetric hyperbolic system \begin{equation*} \dot C =
-2\partial^j C_j, \quad \dot C_j = -\frac12 \partial_j C.
\end{equation*} Thus if $C$ and $C_j$ vanish at $t=0$, they vanish for
all time.

The linearized EC formulation provides an
alternate approach to obtaining a solution of \eqref{G}--\eqref{Cj}
with the given initial data, based on solving a system with better
hyperbolicity properties.  If $g_{ij}$, $K_{ij}$ solve
\eqref{G}--\eqref{Cj}, define
\begin{equation}\label{deff}
f_{kij}= \frac12 [\partial_k g_{ij}-
(\partial^l g_{li}-\partial_i g_l^l)\delta_{jk}-
(\partial^l g_{lj}-\partial_j g_l^l)\delta_{ik}].
\end{equation}
Then $-\partial^kf_{kij}$ coincides with the first three terms of the
right-hand side of \eqref{KADM}, so
\begin{equation}\label{K}
\dot K_{ij}=-\partial^kf_{kij}-\partial_i\partial_j\alpha.
\end{equation}
Differentiating \eqref{deff} in time, substituting \eqref{G},
and using the constraint equation \eqref{Cj}, we obtain
\begin{equation}\label{F}
\dot f_{kij}=-\partial_kK_{ij}+L_{kij},
\end{equation}
where
\begin{equation}\label{defL}
L_{kij}=\partial_k\partial_{(i}\beta_{j)}-
\partial^l\partial_{[l}\beta_{i]}\delta_{jk}-
\partial^l\partial_{[l}\beta_{j]}\delta_{ik}
\end{equation}
The evolution equations \eqref{K} and \eqref{F} for $K_{ij}$
and $f_{kij}$,
together with the evolution equation \eqref{G} for $g_{ij}$,
form the linearized EC system.
As initial data for this system we use the given initial
values of $g_{ij}$ and $K_{ij}$ and derive the initial
values for $f_{kij}$ from those of $g_{ij}$ based on \eqref{deff}:
\begin{equation}\label{finit}
f_{kij}(0)= \frac12 \{\partial_k g_{ij}(0)-
[\partial^l g_{li}(0)-\partial_i g_l^l(0)]\delta_{jk}-
[\partial^l g_{lj}(0)-\partial_j g_l^l(0)]\delta_{ik}\}.
\end{equation}

In this paper we study the preservation of constraints by the
linearized EC  system and the closely related question of the equivalence
of that system and the linearized ADM system. Our main interest is in the
case when the spatial domain is bounded and appropriate boundary
conditions are imposed, but first we consider the result for the pure
Cauchy problem in the remainder of this section.

Suppose that $K_{ij}$ and $f_{kij}$ satisfy the evolution equations
\eqref{K} and \eqref{F} (which decouple from \eqref{G}).
If $K_{ij}$ satisfies the momentum constraint \eqref{Cj} for all
time, then from \eqref{K} we obtain a constraint which must be
satisfied by $f_{kij}$:
\begin{equation}\label{Fc}
\partial^k(\partial^lf_{klj}-\partial_jf_{kl}^{\hphantom{kl}l})=0.
\end{equation}
Note that \eqref{deff} is another constraint that must be satisfied
for all time.
The following theorem shows that the constraints \eqref{Cj}, \eqref{deff}, and 
\eqref{Fc} are preserved by the linearized EC evolution.
\begin{thm}\label{thm:constraint0}
Let initial data $g_{ij}(0)$ and $K_{ij}(0)$ be given satisfying the
constraints \eqref{C} and \eqref{Cj}, respectively, and
$f_{kij}(0)$ be defined by \eqref{finit}.  
Then the unique solution
of the evolution equations \eqref{G}, \eqref{K}, and \eqref{F} satisfies
\eqref{Cj}, \eqref{deff}, and \eqref{Fc} for all time.
\end{thm}
\begin{proof}
First we show that the initial data $f_{kij}(0)$ defined in
\eqref{finit} satisfies the constraint \eqref{Fc}. Applying the
constraint operator in \eqref{Fc} to \eqref{finit} we find
\begin{equation*}
\partial^k(\partial^lf_{klj} -\partial_j
f_{kl}^{\hphantom{kl}l})(0)=
\frac12\partial_j(\partial^l\partial^kg_{kl}-
\partial^k\partial_kg_l^l)(0)=\frac12\partial_jC(0)=0.\quad \mbox{(by \eqref{C})}
\end{equation*}
It is immediate from the evolution equations
that each component $K_{ij}$ satisfies the
inhomogeneous wave equation
\begin{equation*}
\ddot K_{ij}=\partial^k\partial_k K_{ij}-\partial^kL_{kij}-
\partial_i\partial_j\dot\alpha .
\end{equation*}
Applying the momentum constraint operator defined in \eqref{Cj}, we
see that each component $C_j$ satisfies the homogeneous wave equation
\begin{equation}\label{w}
\ddot C_j =\partial^k\partial_k C_j.
\end{equation}
Now $C_j=0$ at the initial time by assumption, so if we can show
that $\dot C_j=0$ at the initial time, we can conclude that $C_j$
vanishes for all time.  But, from \eqref{K} and the definition of $C_j$,
\begin{equation}\label{Cjd}
\dot C_j = -\partial^k(\partial^lf_{klj} -\partial_j
f_{kl}^{\hphantom{kl}l}),
\end{equation}
which we just proved that vanishes at the initial time.  Thus we have shown
$C_j$ vanishes for all time, i.e., \eqref{Cj} holds.  In view
of \eqref{Cjd}, \eqref{Fc} holds as well.
From \eqref{F} and \eqref{G}
we have 
\begin{equation*}
\dot f_{kij}=\frac12 \partial_k\dot g_{ij}-
\partial^l\partial_{[l}\beta_{i]}\delta_{jk}-
\partial^l\partial_{[l}\beta_{j]}\delta_{ik}.
\end{equation*}
Applying the momentum constraint operator to
\eqref{G} and using \eqref{Cj}, it follows that
\begin{equation*}
\frac12 (\partial^l\dot g_{li}-
\partial_i\dot g_l^l)=\partial^l\partial_{[l}\beta_{i]},
\end{equation*}
so
$f_{kij}-[\partial_k g_{ij}-(\partial^lg_{li}-
\partial_ig_l^l)\delta_{kj}-(\partial^lg_{lj}-
\partial_jg_l^l)\delta_{ki}]/2$ does not depend on time.  
From \eqref{finit}, we have \eqref{deff}.
\end{proof}
In view of this theorem it is straightforward to establish the
key result that for given initial data satisfying the
constraints, the unique solution of the linearized EC evolution
equations satisfies the linearized ADM system, and so the linearized
ADM system and the linearized EC system are equivalent.

\begin{thm}\label{thm:equiv0}
Suppose that initial data $g_{ij}(0)$ and $K_{ij}(0)$ are given
satisfying the Hamiltonian constraint \eqref{C} and momentum constraint
\eqref{Cj}, respectively, and that initial data $f_{kij}(0)$ is defined
by \eqref{finit}.  Then the unique solution of the linearized EC
evolution equations \eqref{G}, \eqref{K}, \eqref{F} satisfies the
linearized ADM system \eqref{G}--\eqref{Cj}.
\end{thm}
\begin{proof} 
From Theorem~\ref{thm:constraint0}, we know that $C_j=0$
for all time, i.e., \eqref{Cj} holds.
Then from \eqref{G} and \eqref{Cj} we see that $\dot C=-2\partial^j
C_j=0$, and, since $C$ vanishes at initial time by assumption,
$C$ vanishes for all time, i.e., \eqref{C} holds as well.

It remains to verify \eqref{KADM}. 
From Theorem~\ref{thm:constraint0}, we also have \eqref{deff}.
Substituting \eqref{deff} in \eqref{K} gives \eqref{KADM},
as desired.
\end{proof}

\section{Maximal Nonnegative Constraint Preserving Boundary Conditions}
\label{MNCPBC}
In this section of the paper, we provide maximal nonnegative
boundary conditions for the linearized EC system which are
constraint-preserving in the sense that the analogue of 
Theorem~\ref{thm:constraint0} is true for the initial--boundary value
problem.  This will then imply the analogue of
Theorem~\ref{thm:equiv0}. We assume that $\Omega$ is a polyhedral domain.

Consider an arbitrary face of $\partial\Omega$ and let $n^i$ denote its
exterior unit normal.  Denote by $m^i$ and $l^i$ two additional vectors
which together $n^i$ form an orthonormal basis.  The projection
operator orthogonal to $n^i$ is then given by $\tau_i^j:=m_im^j+l_il^j$
(and does not depend on the particular choice of these tangential
vectors).  Note that
\begin{equation}\label{delid}
\delta_i^j=n_in^j + \tau_i^j, \quad \tau_i^j\tau_j^k=\tau_i^k.
\end{equation}
Consequently,
\begin{equation}\label{dotid}
v_lw^l = n^jv_j n_iw^i + \tau_l^j v_j \tau_i^lw^i \text{\quad for all
$v_l$, $w^l$}. 
\end{equation}

First we consider the following boundary conditions on the face:
\begin{equation}\label{set2a}
n^im^jK_{ij}=n^il^jK_{ij}=n^kn^in^jf_{kij}=n^km^im^jf_{kij}=n^kl^il^jf_{kij}=
n^km^il^jf_{kij}=0.
\end{equation}
These can be written as well:
\begin{equation}\label{set2b}
n^i\tau^{jk}K_{ij}=0, \quad
n^kn^in^jf_{kij}=0, \quad
n^k\tau^{il}\tau^{jm}f_{kij}=0,
\end{equation}
and so do not depend on the choice of basis for the tangent space.
We begin by showing that these boundary conditions are maximal nonnegative
for the hyperbolic system \eqref{G}, \eqref{K}, and \eqref{F}, and so, according to
the classical theory of \cite{F} and \cite{LP} (also \cite{GKO}, \cite{KL},
\cite{MO}, \cite{R}, \cite{Se1}, \cite{Se2}, among others), the initial--boundary 
value problem is well-posed. For convenience, in Appendix A we recall the definition 
and a classical result due to Rauch \cite{R} on $L^2$ well--posedness of maximal 
nonnegative boundary conditions.

Let $V$ denote the vector space of triplets of constant tensors
$(g_{ij},K_{ij},f_{kij})$ all three symmetric with respect to the 
indices $i$ and $j$.  Thus $\dim V=30$.
The boundary operator $A_n$ associated to the evolution equations
\eqref{G}, \eqref{K}, and \eqref{F} is the symmetric linear operator $V\to V$ given by
\begin{equation}\label{An}
\tilde g_{ij}=0,\quad
\tilde K_{ij}=n^kf_{kij}, \quad 
\tilde f_{kij}=n_kK_{ij}.
\end{equation}
A subspace $N$ of $V$ is called nonnegative for $A_n$ if
\begin{equation}\label{ip}
g_{ij}\tilde g^{ij} +K_{ij}\tilde K^{ij} + f_{kij}\tilde f^{kij} \ge 0
\end{equation}
whenever $(g_{ij},K_{ij},f_{kij})\in N$ and $(\tilde g_{ij},\tilde K_{ij},\tilde f_{kij})$
is defined by \eqref{An}.  The subspace is maximal nonnegative if
also no larger subspace has this property.  Since $A_n$ has six
positive, 18 zero, and six negative eigenvalues, a nonnegative subspace
is maximal nonnegative if and only if it has dimension $24$.  Our
claim is that the subspace $N$ defined by \eqref{set2a} is maximal
nonnegative.  The dimension is clearly $24$.  In view of \eqref{An},
the verification of \eqref{ip} reduces to showing that
$n^kf_{kij}K^{ij}\ge0$
whenever \eqref{set2a} holds.  In fact, $n^kf_{kij}K^{ij}=0$,
that is, $n^kf_{kij}$ and $K_{ij}$ are orthogonal (when \eqref{set2a}
holds).
To see this, we use orthogonal expansions of each based
on the normal and tangential components:
\begin{align}\label{expand1}
K_{ij}&=n^ln_in^mn_jK_{lm} + n^ln_i\tau_j^mK_{lm}
+\tau_i^ln^mn_jK_{lm}+\tau_i^l\tau_j^mK_{lm},
\\\label{expand2}
n^kf_{kij}&=n^ln_in^mn_jn^kf_{klm} + n^ln_i\tau_j^mn^kf_{klm}
+\tau_i^ln^mn_jn^kf_{klm}+\tau_i^l\tau_j^mn^kf_{klm}.
\end{align}
In view of the boundary conditions (in the form \eqref{set2b}),
the two inner terms on the right-hand side of \eqref{expand1}
and the two outer terms on the right-hand side of \eqref{expand2}
vanish, 
and so the orthogonality is evident.

Next we show that the boundary conditions are constraint-preserving.
This is based on the following lemma.
\begin{lem}\label{lem}
Suppose that $\alpha$ and $\beta^i$ vanish.
Let $g_{ij}$, $K_{ij}$, and $f_{kij}$ be a solution to the homogeneous hyperbolic system
\eqref{G}, \eqref{K}, and \eqref{F} and suppose that the boundary conditions \eqref{set2a}
are satisfied on some face of $\partial\Omega$.  Let $C_j$ be defined by
\eqref{Cj}.  Then 
\begin{equation}\label{bndid}
\dot C_j n^l\partial_l C^j=0
\end{equation}
on the face.
\end{lem}
\begin{proof}
 In fact we shall
show that $n^jC_j=0$ (so also $n^j\dot C_j=0$) and $\tau_j^p n^l\partial_l C^j=0$, which,
by \eqref{dotid} implies \eqref{bndid}.  First note that
\begin{equation*}
C_j=(\delta_j^m\delta^{ik}-\delta_j^k\delta^{im})\partial_k K_{im}
=(\delta_j^mn^in^k+\delta_j^m\tau^{ik}-\delta_j^k\delta^{im})\partial_k
K_{im},
\end{equation*}
where we have used the first identity in \eqref{delid}.  Contracting
with $n^j$ gives
\begin{align*}
n^jC_j&=(n^mn^in^k+n^m\tau^{ik}-n^k\delta^{im})\partial_k K_{im}
\\
&=-n^mn^in^k\dot f_{kim}+\tau^{il}\tau_l^kn^m\partial_k K_{im}
+n^k \delta^{im}\dot f_{kim}
\end{align*}
where now we have used the equation \eqref{F} (with $\beta_i=0$)
for the first and last term
and the second identity in \eqref{delid} for the middle term.
From the boundary conditions we know that
$n^mn^in^kf_{kim}=0$, and so the first term on the right-hand side
vanishes.  Similarly, we know that $\tau^{il}n^mK_{im}=0$ on the
boundary face, and so the second term vanishes as well (since the
differential operator $\tau_l^k\partial_k$ is purely tangential).
Finally, $n^k\delta^{im}f_{kim}=n^k(n^in^m+l^il^m+m^im^m)f_{kim}=0$,
and so the third term vanishes.
We have established that $n^jC_j=0$ holds on the face.

To show that $\tau_j^p n^l\partial_l C^j=0$ on the face, we start with the identity
\begin{equation*}
\tau_j^p n^l\delta^{mj}\delta^{ik}=\tau^{pm}(n^in^k+\tau^{ik})n^l
=\tau^{pm}n^i(\delta^{kl}-\tau^{kl})+\tau^{pm}\tau^{ik}n^l.
\end{equation*}
Similarly
\begin{equation*}
\tau_j^pn^l\delta^{kj}\delta^{im} =
\tau^{pk}n^ln^in^m+\tau^{pk}\tau^{im}n^l.
\end{equation*}
Therefore,
\begin{align*}
\tau_j^p n^l\partial_l C^j
&=\tau^p_jn^l
\partial_l(\delta^{mj}\delta^{ik}-\delta^{kj}\delta^{im})\partial_k K_{im}
\\
&=(\tau^{pm}n^i\delta^{kl}-\tau^{pm}n^i\tau^{kl}+\tau^{pm}\tau^{ik}n^l
-\tau^{pk}n^ln^in^m-\tau^{pk}\tau^{im}n^l)\partial_k\partial_l K_{im}.
%&=(\tau^{pm}n^i\delta^{kl}-\tau^{pm}n^i\tau^{kl}-\tau^{pk}n^ln^in^m)\partial_l
%\partial_k K_{im}.
\end{align*}
For the last three terms, we again use \eqref{F} to replace
$\partial_l K_{im}$ with $-\dot f_{lim}$ and argue as before to
see that these terms vanish.  For the first term we notice
that $\delta^{kl}\partial_k\partial_l K_{im}=\partial^k\partial_kK_{im}
=\ddot K_{im}$ (from \eqref{K} and \eqref{F} with vanishing
$\alpha$ and $\beta^i$).  Since
$\tau^{pm}n^iK_{im}$ vanishes on the boundary, this term vanishes.
Finally we recognize that the second term is the tangential Laplacian,
$\tau^{kl}\partial_k\partial_l$ applied to the quantity
$n^i\tau^{pm} K_{im}$, which vanishes.  This concludes the proof of
\eqref{bndid}.
\end{proof}
The next theorem asserts that the boundary conditions are
constraint-preserving.
\begin{thm}\label{thm:constraint}
Let $\Omega$ be a polyhedral domain. Given $g_{ij}(0)$ and  
$K_{ij}(0)$ on $\Omega$ satisfying the constraints
\eqref{C} and \eqref{Cj}, respectively, and $f_{kij}(0)$ defined
by \eqref{finit}, define $g_{ij}$, $K_{ij}$, and $f_{kij}$ for
positive time by the evolution equations \eqref{G}, \eqref{K}, and \eqref{F}
and the boundary conditions \eqref{set2a}.  Then the constraints
\eqref{Cj}, \eqref{deff}, and \eqref{Fc} are satisfied for all time.
\end{thm}
\begin{proof}
Exactly as for Theorem~\ref{thm:constraint0} we find that $C_j$
satisfies the wave equation \eqref{w} and both $C_j$ and $\dot C_j$
vanish at the initial time; these facts are unrelated to the
boundary conditions. Define the usual energy
\begin{equation*}
E(t)=\frac12 \int_{\Omega}(\dot C_j\dot
C^j+\partial^lC_j\partial_lC^j)\,dx.
\end{equation*}
Clearly $E(0)=0$.  From \eqref{w} and integration by parts
\begin{equation}\label{dtE}
\dot E=\int_{\partial\Omega}\dot C_jn^l\partial_lC^j d\sigma.
\end{equation}
Therefore, if $\alpha=0$ and $\beta^i=0$, we can invoke
Lemma~\ref{lem}, and conclude that $E$ is constant in time.
Hence $E$ vanishes identically.  Thus $C_j$ is constant, and, since
it vanishes at time $0$, it vanishes for all time. By \eqref{Cjd},
the constraints \eqref{Fc} are also satisfied for all time.
This establishes that the constraints \eqref{Cj} and \eqref{Fc} hold
under the additional assumption that $\alpha$ and $\beta^i$ vanish.

To extend to the case of general $\alpha$ and $\beta^i$ we use
Duhamel's principle.  Let $S(t)$ denote the solution operator
associated to the homogeneous boundary value problem.  That is,
given functions $h_{ij}(0)$, $\kappa_{ij}(0)$, $\phi_{kij}(0)$ on $\Omega$,
define $$S(t)(h_{ij}(0),\kappa_{ij}(0),\phi_{kij}(0))=
(h_{ij}(t),\kappa_{ij}(t),\phi_{kij}(t)),$$ where $h_{ij}$, $\kappa_{ij}$, $\phi_{kij}$ is
the solution to the homogeneous evolution equations
\begin{equation*}
\dot h_{ij}=-2\kappa_{ij},\quad
\dot\kappa_{ij}=-\partial^k\phi_{kij},
\quad \dot\phi_{kij}=-\partial_k\kappa_{ij},
\end{equation*}
satisfying the boundary conditions and assuming the given initial
values.  Then Duhamel's principle represents the solution $g_{ij}$, 
$K_{ij}$, $f_{kij}$ of the inhomogeneous initial-boundary value
problem \eqref{G}, \eqref{K}, \eqref{F}, \eqref{set2a} as
\begin{multline}\label{duhamel}
(g_{ij}(t), K_{ij}(t),f_{kij}(t))=S(t)(g_{ij}(0),K_{ij}(0),f_{kij}(0))\\
+\int_0^t S(t-s)(2\partial_{(i}\beta_{j)},-\partial_i\partial_j\alpha(s),L_{kij}(s))\,ds.
\end{multline}
Now it is easy to check that the Hamiltonian constraint \eqref{C} is satisfied
when $g_{ij}$ is replaced by $2\partial_{(i}\beta_{j)}$ (for any smooth vector
function $\beta^i$), the momentum constraint \eqref{Cj} is
satisfied when $K_{ij}$ is replaced by $-\partial_i\partial_j\alpha(s)$
(for any smooth function $\alpha$), and the constraint \eqref{Fc}
is satisfied when $f_{kij}$ is replaced by $L_{kij}(s)$ defined by 
\eqref{defL} (for any smooth vector function $\beta^i$).  Hence the integrand
in \eqref{duhamel} satisfies the constraints by the result for the homogeneous case,
as does the first term on the right-hand side, and thus the constraints \eqref{Cj} and \eqref{Fc}
are indeed satisfied by $K_{ij}$ and $f_{kij}$, respectively.

The proof of the fact that the constraints \eqref{deff} are satisfied for all time
follows exactly as in Theorem~\ref{thm:constraint0}. 

Note that the boundary conditions \eqref{set2a} play a crucial role in proving that the 
momentum constraints \eqref{Cj} are preserved for all time; 
the preservation of the constraints \eqref{deff} and \eqref{Fc} being
a consequence of this fact.
\end{proof}
The analogue of Theorem~\ref{thm:equiv0} for the initial--boundary
value problem follows from the preceding theorem exactly as before.
\begin{thm}\label{thm:equiv}
Let $\Omega$ be a polyhedral domain.
Suppose that initial data $g_{ij}(0)$ and $K_{ij}(0)$ are given
satisfying the Hamiltonian constraint \eqref{C} and momentum constraint
\eqref{Cj}, respectively, and that initial data $f_{kij}(0)$ is defined
by \eqref{finit}.  Then the unique solution of the linearized EC
initial--boundary value problem \eqref{G}, \eqref{K}, \eqref{F},
together with the boundary conditions \eqref{set2a}
satisfies the linearized ADM system
\eqref{G}--\eqref{Cj} in $\Omega$.
\end{thm}

We close this section by noting a second set of boundary
conditions which are maximal nonnegative and constraint-preserving.
These are
\begin{equation}\label{set1a}
n^in^jK_{ij}=m^im^jK_{ij}=l^il^jK_{ij}=m^il^jK_{ij}=n^kn^im^jf_{kij}=
n^kn^il^jf_{kij}=0,
\end{equation}
or, equivalently,
\begin{equation}\label{set1b}
n^in^jK_{ij}=0, \quad
\tau^{il}\tau^{jm}K_{ij}=0, \quad
n^kn^i\tau^{jl}f_{kij}=0.
\end{equation}
Now when we make an orthogonal expansion as in \eqref{expand1},
\eqref{expand2}, the outer terms on the right-hand side of the
first equation and the inner terms on the right-hand side of the
second equation vanish (it was the reverse before), so we
again have the necessary orthogonality to demonstrate that
the boundary conditions are maximal nonnegative.  Similarly, to
prove the analogue of Lemma~\ref{lem}, for these boundary
conditions we show that the tangential component of $\dot C_j$ vanishes
and the normal component of $n^l\partial_l C^j$ vanishes (it was
the reverse before).  Otherwise the analysis is essentially the
same as for the boundary conditions \eqref{set2a}.

\section{Extended EC System}
\label{EECS}
In this section we indicate an extended initial boundary value
problem whose solution solves the linearized ADM system 
\eqref{G}--\eqref{Cj} in $\Omega$. This approach could present advantages
from the numerical point of view since the momentum constraint
is ``built-in,'' and so controlled for all time.
The new system consists of \eqref{G}, \eqref{F},
and two new sets of equations corresponding to \eqref{K} 
\begin{equation}\label{KE}
\dot{K}_{ij}=-\partial^kf_{kij}+\frac{1}{2} (\partial_ip_j+\partial_jp_i)-
\partial^kp_k\delta_{ij}-\partial_i\partial_j\alpha,
\end{equation}
and to
a new three dimensional vector field $p_i$ defined by
\begin{equation}\label{p}
\dot{p}_i=\partial^lK_{li}-\partial_iK_l^l.
\end{equation}
Observe that the additional terms that appear on the right-hand side
of \eqref{KE} compared with \eqref{K} are nothing but the negative components
of the formal adjoint of the momentum constraint operator applied to $p_i$.

Let $\tilde{V}$ be the vector space of quadruples of constant tensors
$(g_{ij},K_{ij},f_{kij},p_k)$ symmetric with respect to the indices
$i$ and $j$. Thus $\dim \tilde{V}=33$. The boundary operator 
$\tilde{A}_n :\tilde{V}\to\tilde{V}$ in this case is given by
\begin{equation}\label{Andef}
\tilde{g}_{ij}=0,\ 
\tilde{K}_{ij}=n^kf_{kij}-\frac{1}{2}(n_ip_j+n_jp_i)+n^kp_k\delta_{ij},
\ 
\tilde{f}_{kij}=n_kK_{ij},\ 
\tilde{p}_i=-n^lK_{il}+n_iK_l^l.
\end{equation}
The boundary operator $\tilde{A}_n$ associated to the evolution equations
\eqref{G}, \eqref{KE}, \eqref{F}, and \eqref{p} has six positive, 21
zero, and six negative eigenvalues. Therefore, a nonnegative subspace
is maximal nonnegative if and only if it has
dimension 27. We claim that the following boundary conditions are
maximal nonnegative for \eqref{G}, \eqref{KE}, \eqref{F}, and \eqref{p}
\begin{equation}\label{set2ae}
\begin{gathered}
n^im^jK_{ij}=n^il^jK_{ij}=n^kn^in^jf_{kij}=n^k(m^im^jf_{kij}+p_k)=\\
n^k(l^il^jf_{kij}+p_k)=n^km^il^jf_{kij}=0.
\end{gathered}
\end{equation}
These can be written as well:
\begin{equation}\label{set2be}
n^i\tau^{jk}K_{ij}=0, \quad
n^kn^in^jf_{kij}=0, \quad
n^k(\tau^{il}\tau^{jm}f_{kij}+\tau^{lm}p_k)=0,
\end{equation}
and so do not depend on the choice of basis for the tangent space.

Let us prove the claim that the subspace $\tilde{N}$ defined by
\eqref{set2ae} is maximal nonnegative. Obviously,
$\dim\tilde{N}=27$. Hence, it remains to be proven that $\tilde{N}$ is
also nonnegative. In view of \eqref{Andef}, the verification of
non-negativity of $\tilde{N}$ reduces to showing that
\begin{equation}\label{positivity}
n^kf_{kij}K^{ij}-n^ip^jK_{ij}+n^kp_kK_l^l\geq 0
\end{equation}
whenever \eqref{set2ae} holds. In fact, we can prove that the
left-hand side of \eqref{positivity} vanishes pending 
\eqref{set2ae} holds. From the boundary conditions (in the
form \eqref{set2be}) and the orthogonal expansions \eqref{expand1} and
\eqref{expand2} of $K_{ij}$ and $f_{kij}$, respectively, the first
term on the right-hand side of \eqref{positivity} reduces to
$n^k\tau^{il}\tau^{jm}f_{kij}K_{lm}=-n^kp_k\tau^{lm}K_{lm}$.
Then, combining the first and third terms of the left-hand side of 
\eqref{positivity} gives
$-n^kp_k\tau^{ij}K_{ij}+n^kp_k\delta^{ij}K_{ij}=n^kp_kn^in^jK_{ij}$.
Finally, by using the orthogonal decomposition
$p^j=n^kp_kn^j+\tau^{kj}p_k$ and the first part of the boundary 
conditions \eqref{set2be}
the second term of the left-hand side of \eqref{positivity} is
$-n^kp_kn^in^jK_{ij}-p_kn^i\tau^{kj}K_{ij}=-n^kp_kn^in^jK_{ij}$, which
is precisely the negative sum of the first and third terms of the
left-hand side of \eqref{positivity}. This concludes the proof of 
\eqref{positivity}.  

\begin{thm}\label{thm:extended}
Let $\Omega$ be a polyhedral domain. Suppose that the initial data
$g_{ij}(0)$ and $K_{ij}(0)$ are given satisfying the Hamiltonian \eqref{C}
and momentum constraints \eqref{Cj}, respectively, $f_{kij}(0)$ is 
defined by \eqref{deff}, and $p_i(0)=0$. Then the unique solution
$(g_{ij},K_{ij},f_{kij},p_i)$ of the initial boundary value problem
\eqref{G}, \eqref{KE}, \eqref{F}, and \eqref{p}, together with the
boundary conditions \eqref{set2ae}, satisfies the properties
$p_i=0$ for all time, and $(g_{ij},K_{ij})$ solves the linearized
ADM system \eqref{G}--\eqref{Cj} in $\Omega$.
\end{thm}
\begin{proof}
Observe that the solution of the initial boundary value problem
\eqref{G}, \eqref{K}, \eqref{F}, and \eqref{set2a} (boundary
conditions), together with $p_i=0$ for all time, is the unique
solution of the initial boundary value problem \eqref{G}, \eqref{KE}, 
\eqref{F}, and \eqref{p}, together with the boundary conditions \eqref{set2ae}.
The conclusion follows from Theorem~\ref{thm:equiv}.
\end{proof}
We close by indicating a second set of maximal nonnegative boundary
conditions (corresponding to \eqref{set1a}) for \eqref{G}, \eqref{KE}, 
\eqref{F}, and \eqref{p} for which Theorem~\ref{thm:extended} holds as
well. These are 
\begin{equation}\label{set1ae}
\begin{gathered}
n^in^jK_{ij}=m^im^jK_{ij}=m^il^jK_{ij}=l^il^jK_{ij}=\\
n^kn^im^jf_{kij}-m^kp_k=
n^kn^il^jf_{kij}-l^kp_k=0,
\end{gathered}
\end{equation}
or, equivalently,
\begin{equation}\label{set1be}
n^in^jK_{ij}=0, \quad
\tau^{il}\tau^{jm}K_{ij}=0, \quad
n^kn^i\tau^{jl}f_{kij}-\tau^{kl}p_k=0.
\end{equation}

\section{Inhomogeneous Boundary Conditions}
\label{IBC}
In this section we provide a formal method of constructing well-posed constraint-preserving
{\it inhomogeneous} boundary conditions for \eqref{G}, \eqref{K}, and \eqref{F}
corresponding to the two sets of boundary conditions \eqref{set2a} and
\eqref{set1a}, respectively. The first set of inhomogeneous boundary
conditions corresponds to \eqref{set2a} and can be written in the
following form
\begin{equation}\label{set2ai}
n^im^j\tilde{K}_{ij}=n^il^j\tilde{K}_{ij}=n^kn^in^j\tilde{f}_{kij}=n^km^im^j\tilde{f}_{kij}=
n^kl^il^j\tilde{f}_{kij}=
n^km^il^j\tilde{f}_{kij}=0,
\end{equation}
where $\tilde{K}_{ij}=K_{ij}-\kappa_{ij}$,
$\tilde{f}_{kij}=f_{kij}-F_{kij}$, with $\kappa_{ij}$ and $F_{kij}$
given in $\overline{\Omega}$ for all time and satisfying the constraints \eqref{Cj} and \eqref{Fc},
respectively.

The analogue of Theorem~\ref{thm:constraint} for the inhomogeneous boundary
conditions \eqref{set2ai} is true.

\begin{thm}\label{thm:constrainti}
Let $\Omega$ be a polyhedral domain. Given $g_{ij}(0)$ and  
$K_{ij}(0)$ on $\Omega$ satisfying the constraints
\eqref{C} and \eqref{Cj}, respectively, and $f_{kij}(0)$ defined
by \eqref{finit}, define $g_{ij}$, $K_{ij}$, and $f_{kij}$ for
positive time by the evolution equations \eqref{G}, \eqref{K}, and \eqref{F}
and the boundary conditions \eqref{set2ai}.  Then the constraints
\eqref{Cj}, \eqref{deff}, and \eqref{Fc} are satisfied for all time.
\end{thm}
\begin{proof}
Observe that $g_{ij}$, $\tilde{K}_{ij}$, and $\tilde{f}_{kij}$ satisfy 
\eqref{G}, \eqref{K}, and \eqref{F} with the forcing terms replaced by
$2\partial_{(i}\beta_{j)}$, $-\partial_i\partial_j\alpha-\partial^kF_{kij}-\dot{\kappa}_{ij}$, and
$L_{kij}-\partial_k\kappa_{ij}-\dot{F}_{kij}$, respectively.
Exactly as in Theorem~\ref{thm:constraint}, it follows that
$\tilde{K}_{ij}$ and $\tilde{f}_{kij}$ satisfy \eqref{Cj}
and \eqref{Fc}, respectively, for all time. Thus, $K_{ij}$ and
$f_{kij}$ satisfy \eqref{Cj} and \eqref{Fc}, respectively, for all time.
Finally, same arguments as in Theorem~\ref{thm:constraint0} show that the
constraints \eqref{deff} are also preserved through evolution for all time.
\end{proof}
The analogue of Theorem~\ref{thm:equiv} for the case of the
inhomogeneous boundary conditions \eqref{set2ai} follows from
the preceding theorem by using the same arguments as in the proof
of Theorem~\ref{thm:equiv0}.

\begin{thm}\label{thm:equivi}
Let $\Omega$ be a polyhedral domain.
Suppose that initial data $g_{ij}(0)$ and $K_{ij}(0)$ are given
satisfying the Hamiltonian constraint \eqref{C} and momentum constraint
\eqref{Cj}, respectively, and that initial data $f_{kij}(0)$ is defined
by \eqref{finit}.  Then the unique solution of the linearized EC
initial--boundary value problem \eqref{G}, \eqref{K}, \eqref{F},
together with the {\it inhomogeneous} boundary conditions \eqref{set2ai}
satisfies the linearized ADM system
\eqref{G}--\eqref{Cj} in $\Omega$.
\end{thm}

Note that there is a second set of inhomogeneous
boundary conditions corresponding to \eqref{set1a} for which 
Theorem~\ref{thm:constrainti} and Theorem~\ref{thm:equivi} remain
valid. These are
\begin{equation}\label{set1ai}
n^in^j\tilde{K}_{ij}=m^im^j\tilde{K}_{ij}=l^il^j\tilde{K}_{ij}=
m^il^j\tilde{K}_{ij}=n^kn^im^j\tilde{f}_{kij}=n^kn^il^j\tilde{f}_{kij}=0,
\end{equation}
where again $\tilde{K}_{ij}=K_{ij}-\kappa_{ij}$,
$\tilde{f}_{kij}=f_{kij}-F_{kij}$, with $\kappa_{ij}$ and $F_{kij}$
given and satisfying the constraints \eqref{Cj} and \eqref{Fc},
respectively.

Similar considerations can be made for the extended system introduced
in the previous section. There are two sets of {\it inhomogeneous} 
boundary conditions for which the extended system produces solutions 
of the linearized ADM system \eqref{G}--\eqref{Cj} on a polyhedral domain $\Omega$.
These are
\begin{equation}\label{set2aei}
\begin{gathered}
n^im^j\tilde{K}_{ij}=n^il^j\tilde{K}_{ij}=n^kn^in^j\tilde{f}_{kij}=
n^k(m^im^j\tilde{f}_{kij}+p_k)=n^k(l^il^j\tilde{f}_{kij}+p_k)=\\
n^km^il^j\tilde{f}_{kij}=0
\end{gathered}
\end{equation}
and
\begin{equation}\label{set1aei}
\begin{gathered}
n^in^j\tilde{K}_{ij}=m^im^j\tilde{K}_{ij}=m^il^j\tilde{K}_{ij}=
l^il^j\tilde{K}_{ij}=n^kn^im^j\tilde{f}_{kij}-m^kp_k=\\
n^kn^il^j\tilde{f}_{kij}-l^kp_k=0,
\end{gathered}
\end{equation}
where $\tilde{K}_{ij}$ and $\tilde{f}_{kij}$ are
defined as before.

The next theorem is an extension of Theorem~\ref{thm:extended} to the
case of inhomogeneous boundary conditions.

\begin{thm}\label{thm:extendedi}
Let $\Omega$ be a polyhedral domain. Suppose that the initial data
$g_{ij}(0)$ and $K_{ij}(0)$ are given satisfying the Hamiltonian \eqref{C}
and momentum constraints \eqref{Cj}, respectively, $f_{kij}(0)$ is 
defined by \eqref{deff}, and $p_i(0)=0$. Then the unique solution
$(g_{ij},K_{ij},f_{kij},p_i)$ of the initial boundary value problem
\eqref{G}, \eqref{KE}, \eqref{F}, and \eqref{p}, together with the
{\it inhomogeneous} boundary conditions \eqref{set2aei} (or
\eqref{set1aei}), satisfies the properties
$p_i=0$ for all time, and $(g_{ij},K_{ij})$ solves the linearized
ADM system \eqref{G}--\eqref{Cj} in $\Omega$.
\end{thm}
\begin{proof}
Note that the solution of the initial boundary value problem
\eqref{G}, \eqref{K}, \eqref{F}, and \eqref{set2ai} (or
\eqref{set1ai}, respectively), together
with $p_i=0$ for all time, is the unique solution of the initial
boundary value problem \eqref{G}, \eqref{KE}, \eqref{F}, and
\eqref{p},
together with the boundary conditions \eqref{set2aei} (or
\eqref{set1aei}, respectively). The conclusion follows from
Theorem~\ref{thm:equivi}.
\end{proof}

\section{Concluding Remarks}\label{CR}
We have studied the boundary conditions problem for the standard 
EC formulation of Einstein's equations linearized 
about the Minkowski spacetime. In Section~\ref{MNCPBC}, we indicate
two sets of maximal nonnegative boundary conditions \eqref{set2a} and
\eqref{set1a}, respectively, which are consistent with the constraints.
These boundary conditions were announced in \cite{AT} and overlap with
the boundary conditions found in \cite{CPRST} for the generalized
EC formulation for $0<\eta<2$ with vanishing shift and lapse density
perturbations. However, the energy method of \cite{CPRST} works only
for the generalized EC formulation with $0<\eta<2$; the standard EC
formulation corresponds to $\eta=4$. Moreover, we prove that our
boundary conditions are well-posed and consistent with the constraints
in the more general case of arbitrary shift and lapse density perturbations
by using a new argument involving the Duhamel's principle. 
Also, our approach emphasizes the relation between the ADM formulation and
the constrained evolution of the EC system in the linearized context. Besides, 
our method is simpler, yet effective, and seems to be easily transferable to other
formulations and/or other background spacetimes. In fact, other Einstein's hyperbolic 
formulations, e.g., Alekseenko--Arnold \cite{AA}, are analyzed in \cite{Ta} 
by using the same method. A subclass of the boundary conditions presented
in this paper and introduced previously in \cite{AT} has been pointed out 
by Frittelli and Gomez in \cite{FG0} (in the case of vanishing shift vector) 
as an example to their Einstein boundary conditions, that is, the vanishing 
of the projection of Einstein's tensor along the normal to the boundary.

One of the main results in this paper is the construction of an extended 
symmetric hyperbolic system which incorporates the momentum constrains 
as main variables. For this extended system, we construct two sets of 
maximal nonnegative boundary conditions and establish its relationship with
the linearized ADM formulation. Such a construction could serve as a model
of how to control the bulk constraint violations by making the constraints
part of the main evolution system, and so keeping them under control for
all time. To the best of our knowledge, this is a new approach regarding
the bulk constraint violations control.

We also make some considerations about how inhomogeneous boundary 
conditions consistent with the constraints could be constructed.

In some places, our methods of proof interfere with 
the techniques used in \cite{CLT} and \cite{CPRST}, e.g., using
the trading of normal derivatives for tangential and temporal ones and
the use of the energy method to prove that the constraints are preserved.
We apply these techniques to the slightly more general case of polyhedral
domains (as opposed to cubic boxes) and in a more systematic way. 
This could be of potential interest to the case of curved boundary domains,
for which the derivative components trading techniques introduce new
terms related to the geometry of the boundaries (see \cite{Ta}, Section 4.2,
for the analysis of a model problem similar to the linearized EC formulation on
curved domains). It is also expected that these or similar techniques will be useful
in the nonlinear case. For the interested reader, we point out the work done
in \cite{KLSBP}, where the authors construct new boundary conditions for the
nonlinear KST form \cite{KST} of the Einstein equations (which includes the EC
formulation).  Their boundary conditions are designed to prevent the 
influx of constraint violations and physical gravitational waves into the 
computational domain. However, as specified in \cite{KLSBP},
there is no rigorous mathematical well-posedness theoretical ground yet for these
kind of boundary conditions, as opposed to the simpler case of maximal nonnegative
boundary conditions.

\medskip

\appendix
\section{Maximal nonnegative boundary conditions for symmetric hyperbolic systems}

Let $\Omega\subset\R^n$ be a bounded domain with smooth boundary and $T>0$. We introduce
the notations $\cO=(0,T)\x\Omega$ and $\Gamma=(0,T)\x\partial\Omega$. Consider the first
order differential operator $L:= \partial_t\,+\sum_{i=1}^nA_i(t,x)\partial_i\, +B(t,x)$,
where $A_i\in$Lip$(\overline{\cO})$, $B\in L^\infty(\overline{\cO})$, and
$(B+B^*)/2-\sum_{i=1}^n\partial_i A_i\in L^\infty (\cO )$.
We suppose that $L$ is symmetric, that is, $A_i=A_i^*$ on $\overline{\cO}$.
Our interest is in solving the initial--boundary value problem
\begin{equation}\label{AppendIVP}
Lu=f(t,x) \ \mbox{in } \cO ,\quad
u(0, \cdot)=g \ \mbox{in } \Omega ,\quad
u(t,x)\in N(t,x) \ \mbox{for }(t,x)\in [0,T]\x\partial\Omega, 
\end{equation}
where $N(t,x)$ is a Lipshitz continuous map from $[0,T]\x\partial\Omega$
to the subspaces of $\C^n$. 
Set $n^i$ be the outer unit normal to $\Gamma$, and denote by $A_n(t,x)$ the
{\it boundary matrix/operator} $A_n(t,x):=\sum_{i=1}^nn^i(x)A_i(t,x)$.
We assume that $\Gamma$ is characteristic of constant multiplicity in the sense
that $\dim \ker A_n$ is constant on each component of $\Gamma$. We next suppose
that $N$ is {\it maximal nonnegative}, that is, the following two conditions hold
on $\overline\Gamma$:
\begin{equation}\label{mp1}
\langle A_n(t,x)v,v\rangle\geq 0,\ \forall (t,x)\in \overline\Gamma,\ \forall v\in N(t,x)
\end{equation}
and
\begin{equation}\label{mp2}
\dim N(t,x)=\#\ \text{nonnegative eigenvalues of }A_n(t,x)\text{ counting multiplicity}.
\end{equation}
The maximality condition \eqref{mp2} implies that the boundary subspace $N$ cannot be 
enlarged while preserving \eqref{mp1}.

Let $\mathbb{H}_{\cO}:=\{ u\in L^2(\cO):\, Lu\in L^2(\cO)\}$. It is easy to prove that
$\mathbb{H}_{\cO}$ is a Hilbert space with respect to the inner product
$\langle u,u\rangle_{\mathbb{H}_{\cO}}:=\langle u,u\rangle_{L^2(\cO)}+\langle Lu,Lu\rangle_{L^2(\cO)}$.
\begin{thm}
{\rm ($L^2$ well-posedness, Theorem 9 in \cite{R})} For any $f\in L^1((0,T):L^2(\Omega))$ and $g\in L^2(\Omega)$
there is a unique $u\in\mathbb{H}_{\cO}$ satisfying \eqref{AppendIVP}. In addition, $u\in C((0,T):L^2(\Omega))$,
\begin{equation*}
\sup_{0\leq t\leq T}\| u(t)\|_{L^2(\Omega)}\leq C\| f\|_{L^1((0,T):L^2(\Omega))}+\| g\|_{L^2(\Omega)},
\end{equation*}
and
\begin{equation*}
\| u(t_2)\|_{L^2(\Omega)}-\| u(t_1)\|_{L^2(\Omega)}\leq 
\int_{t_1}^{t_2}\| f(\sigma)\|_{L^2(\Omega)}+ C\| u(\sigma)\|_{L^2(\Omega)}\, d\sigma .
\end{equation*}
\end{thm}

%\newpage
\end{document}